\documentclass[aps,preprint,amsfonts,amssymb,superscriptaddress]{revtex4}%
\usepackage{amsmath}
\usepackage{graphicx}%
\usepackage{amsfonts}%
\usepackage{amssymb}

\begin{document}

\title{Detailed discussion of a linear electric field frequency shift induced in
confined gases by a magnetic field gradient: Implications for neutron electric
dipole moment experiments.}
\author{S.K. Lamoreaux}
\affiliation{University of California, Los Alamos National Laboratory, Physics Division,
Los Alamos, NM 87545}
\author{R. Golub}
\affiliation{Physics Department, North Carolina State University, Raleigh, N.C. (present
address: Hahn Meitner Institut, Berlin, Germany)}

\begin{abstract}
The search for particle electric dipole moments (EDM) is one of the best
places to look for physics beyond the Standard Model of electroweak
interaction because the size of time reversal violation predicted by the
Standard Model is incompatible with present ideas concerning the creation of
the Baryon-Antibaryon asymmetry. As the sensitivity of these EDM searches
increases more subtle systematic effects become important. We develop a
general analytical approach to describe a systematic effect recently observed
in an electric dipole moment experiment using stored particles \cite{JMP}. Our
approach is based on the relationship between the systematic frequency shift
and the velocity autocorrelation function of the resonating particles. Our
results, when applied to well-known limiting forms of the correlation
function, are in good agreement with both the limiting cases studied in recent
work that employed a numerical/heuristic analysis. Our general approach
explains some of the surprising results observed in that work and displays the
rich behavior of the shift for intermediate frequencies, which has not been
studied previously. In an appendix we give a new derivation of Egelstaf's
theorem which we used in our study of the Diffusion theory (low frequency)
limit of the effect.

\end{abstract}
\volumeyear{year}
\volumenumber{number}
\issuenumber{number}
\eid{identifier}
\date{January 23, 2005}
\startpage{1}
\maketitle
\tableofcontents

\section*{\bigskip Introduction}

The search for an electric dipole moment (EDM) of the neutron is perhaps
unique in modern physics in that experimental work on this subject has been
going on more or less continuously for over 50 years. In that period the
experimental sensitivity has increased by more than a factor of $10^{6}$
without an EDM ever being observed. The reason for this apparently obsessive
behavior by a small group of dedicated physicists is that the observation of a
non-zero neutron EDM would be evidence of time reversal violation and for
physics beyond the so-called Standard Model of electroweak interactions. An
essential point is that the Standard Model predictions of the magnitude of
time reversal violation are inconsistent with our ideas of the formation of
the universe; namely the production of the presently observed matter -
\ anti-matter asymmetry requires time reversal violation many orders of
magnitude greater than that predicted by the Standard Model.

In this type of experiment (null experiment) the control of systematic errors
is of great significance. While the switch of experimental technique from beam
experiments to experiments using stored Ultra-cold neutrons (UCN) has
eliminated many of the sources of systematic error associated with the beam
technique, the gain in sensitivity brought by the new UCN technique means that
the experiments are sensitive to a new range of systematic errors. One of the
most serious of these is associated with the interaction of gradients of the
ever-present constant magnetic field with the well known motional magnetic
field $\left(  \frac{\overrightarrow{v}}{c}\times\overrightarrow{E}\right)  $.
As the particles move in the apparatus, these fields, as seen by the
particles, will be time dependent. This effect was first pointed out by
Commins \cite{commins} and explained in terms of the geometrical phase
concept. A more general description is in terms of the Bloch-Siegert shift of
magnetic resonance frequencies due to the time dependent fields mentioned
above \cite{JMP},\cite{Ramsey}.

The effect was apparently empirically identified in the ILL Hg comagnetometer
EDM experiment and recently Pendlebury et al \cite{JMP} have given a very
detailed discussion of it, including intuitive models and analytical
calculations for certain cases, the relation between and regions of
applicability of the geometric phase and Bloch-Siegert models, numerical
simulations and experimental verification of the most significant features.
However this pioneering work has left certain questions unanswered. In
particular the understanding of effects of collisions on the systematic
frequency \ shifts remains incomplete.

In this work we attempt to clarify several points concerning the influence of
particle collisions. We explain the reason that in contrast to gas collisions,
collisions with the walls were observed to have no effect on the magnitude of
the systematic frequency shifts and show that this only applies to the
limiting cases of high and low frequency. We show that the frequency shift is
related to the velocity autocorrelation function of the resonating particles.
Our solution, when applied to well known limiting forms of the correlation
function, gives results in agreement with those obtained numerically in
\cite{JMP}. McGregor has taken a similar approach to the problem of relaxation
due to static field gradients \cite{McGregor}, whereas the approach taken by
Cates et al to the problem of static field gradients \cite{Cates1} and
gradients combined with oscillating perturbing fields \cite{Cates2} is
somewhat different than ours.

\subsection{Brief description of the effect}

Consider a case where, in a storage experiment, there is a radial magnetic
field due to a magnetic field gradient in the $z$ direction ( $B_{0}$, the
quantization axis, and the electric field $E$ are along $z$). Now consider
roughly circular orbits, due to specular reflection around the bottle at a
constant angle, in the $x-y$ plane with radius approximately the bottle radius
$R$. The wall collisions occur at a frequency $1\//\tau_{c}$ while the orbital
frequency is $\omega_{r}=2\alpha/\tau_{c}$ where $\alpha$ is the incidence
angle relative to the surface. We can transform into a rotating frame at
$\omega_{r}$ (note that this is not the Schwinger rotating frame that
eliminates $B_{0}$) so that the problem is quasi-static \cite{Ramsey}.

The radial field, with the barrel gradient plus $v\times E$ field, is
\[
B_{R}=B_{r}\pm B_{E}=aR\pm{\frac{\omega_{r}RE}{c}}%
\]
where $B_{r}(r)=\left(  r/2\right)  \partial B_{z}/\partial z=ar$ is the
radial field due to the axial gradient, and $\pm B_{E}=r\omega_{r}E/c$ is the
radially directed $v\times E$ field and the $\pm$ refer to the rotation direction.

In the rotating frame,
\[
B^{2}=(B_{0}-\omega_{r}/\gamma)^{2}+(B_{R})^{2}%
\]
where $\gamma$ is the gyromagnetic ratio. Expanding in the limit where
$B_{R}<<B_{0}$ with transformation back to the lab frame we find
\[
B=B_{0}+{\frac{1}{2}}{\frac{(aR-\omega_{r}RE/c)^{2}}{B_{0}-\omega_{r}/\gamma}%
}=B_{0}-{\frac{(aR^{2}\omega_{r}E/c)}{B_{0}-\omega_{r}/\gamma}}%
\]
keeping only terms linear in $B_{E}.$ Averaging over rotation direction (e.g.,
the sign of $\omega_{r}$, the net effect of the gradient field combined with a
$v\times E$ yields a systematic (magnetic field) shift of%
\begin{equation}
\delta\omega=\gamma\delta B=-{\frac{\gamma^{2}av^{2}E}{c\left(  \omega_{0}%
^{2}-\omega_{r}^{2}\right)  }} \label{BS}%
\end{equation}
equivalent to Eq. (18) of \cite{JMP}. Taking the limit $\omega_{r}/\gamma\ll
B_{0}$ we have
\begin{equation}
\delta B=-{\frac{aR^{2}\omega_{r}^{2}E}{\gamma cB_{0}^{2}}},\quad\delta
\omega=-{\frac{aR^{2}\omega_{r}^{2}E}{cB_{0}^{2}}} \label{1}%
\end{equation}
where which would seem to set the scale of the effect and is equivalent to Eq.
(19) of \cite{JMP}. In this limit, the frequency shift does not depend on
$\gamma$, implying that it is the result of a geometric effect.

In the other limit, where the rotation frequency is much faster than the
Larmor frequency, we similarly find that
\begin{equation}
\delta B=\gamma aR^{2}E/c,\quad\delta\omega=\gamma^{2}aR^{2}E/c \label{2}%
\end{equation}
which is independent of the motional frequency $\omega_{r}$ of opposite sign
from the previous limit and equivalent to Eq. (21) of \cite{JMP}.

\section{\bigskip Frequency shift due to fluctuating fields in the x-y plane \ \ }

\subsection{\label{densmatr}Density matrix approach to the problem}

The issues of the effects of a weak fluctuating potential on the evolution of
the density matrix have been well-addressed in the literature. However, these
treatments generally assume that the perturbing potential has a short
correlation time, and certain assumptions regarding averaging are not
applicable to our problem. The effect of a static electric field $E$ by itself
was treated in \cite{lam} where the $E^{2}$ effect was related to the
correlation time, and requirements on the field reversal accuracy were discussed.

So we therefore start from the beginning, following \cite{Abragam} (p. 276).

The radial gradient and $v\times E$ fields can be treated as weak fluctuating
perturbing fields $B_{x,y}(t)$ in the $x-y$ plane, with a constant $B_{0}$
applied along $z$. The perturbing fields $B_{x,y}^{\prime}(t)$ can be written
as
\begin{equation}
B_{x}^{\prime}(t)=B_{x}(t)-\langle B_{x}(t)\rangle;\ \ \ \ \ B_{y}^{\prime
}(t)=B_{y}(t)+\langle B_{y}(t)\rangle
\end{equation}
where $\langle....\rangle$ represents a time average of $B_{x,y}(t)$. The
constant component of the perturbing field are added to $B_{0}$,
\begin{equation}
B_{0}^{\prime}=\sqrt{(B_{0})^{2}+\langle B_{x}(t)\rangle^{2}+\langle
B_{y}(t)\rangle^{2}}%
\end{equation}
leaving the perturbing fields with averages of zero. We define
\begin{equation}
\omega_{0}=\gamma B_{0}^{\prime};\ \ \ \ \omega_{x,y}(t)=\gamma B_{x,y}%
^{\prime}(t).
\end{equation}

The Hamiltonian is thus
\begin{equation}
H=-{\frac{\omega_{0}}{2}}\sigma_{z}-{\frac{\omega_{x}}{2}}\sigma
_{x}-{\frac{\omega_{y}}{2}}\sigma_{y}=H_{0}+H_{1}(t).
\end{equation}

Defining
\begin{equation}
2b=\omega_{x}+i\omega_{y};\ \ \ \ 2b^{\ast}=\omega_{x}-i\omega_{y}%
\end{equation}
the perturbing Hamiltonian can be rewritten as
\begin{equation}
H_{1}(t)=b^{\ast}\sigma_{+}+b\sigma_{-}%
\end{equation}
where $\sigma_{\pm}$ are defined in the appendix, and it is understood that
$b$ is intrinsically time-dependent. Furthermore, the density matrix can be
expanded in the spherical Pauli basis,
\begin{equation}
\rho=1+\rho_{1,0}\sigma_{z}+\rho_{1,1}\sigma_{+}+\rho_{1,-1}\sigma_{-}
\label{densexpan}%
\end{equation}
where $\rho_{11}=\rho_{1-1}^{\ast}$.

The time evolution of the density matrix is
\begin{equation}
{\frac{d\rho}{dt}}=-i[H_{0}+H_{1}(t),\rho].
\end{equation}
The explicit dependence on the constant $H_{0}$ can be eliminated by
transforming to the rotating frame (also called the interaction
representation), with
\begin{equation}
H_{1}(t)\rightarrow e^{iH_{0}t}H_{1}(t)e^{-iH_{0} t};\ \ \ \rho\rightarrow
e^{iH_{0}t}\rho e^{-iH_{0} t}%
\end{equation}
where
\begin{equation}
e^{iH_{0}t}= \left(
\begin{array}
[c]{cc}%
e^{-i\omega_{0}t/2} & 0\\
0 & e^{i\omega_{0} t/2}%
\end{array}
\right)
\end{equation}
We henceforth will work in the rotating frame, with
\begin{equation}
H_{1}(t)=e^{-i\omega_{0} t} b^{\ast}\sigma_{+} + e^{i\omega_{0} t}\sigma_{-}.
\end{equation}

The time evolution of the density matrix in the rotating frame is
\begin{equation}
{\frac{d\rho}{dt}}=-i[H_{1}(t),\rho]
\end{equation}
which can be integrated by successive approximations to
\begin{align}
\rho(t)  &  =\rho(0)-i\int_{0}^{t}[H_{1}(t^{\prime}),\rho(0)]dt^{\prime
}\nonumber\\
&  -\int_{0}^{t}dt^{\prime}\int_{0}^{t^{\prime}}dt^{\prime\prime}%
[H_{1}(t^{\prime}),[H_{1}(t^{\prime\prime}),\rho(0)]].
\end{align}
We are interested in the relaxation rates and frequency shifts due to the
perturbing fields, which can be found through the time derivative of $\rho$,
which by introducing a new variable $\tau=t-t^{\prime\prime}$, is
\begin{equation}
{\frac{d\rho}{dt}}=-i[H_{1}(t),\rho(0)]-\int_{0}^{t}d\tau\lbrack
H_{1}(t),[H_{1}(t-\tau),\rho(0)]].
\end{equation}
The first term on the r.h.s has an ensemble average of zero; furthermore,
there is no correlation between $\rho$ and the fluctuating Hamiltonian (e.g.,
phases of the neutrons have no explicit spatial dependence, and $H_{1}(t)$ is
different for every neutron in the system). In addition, if we assume the
perturbation is weak, $\rho(0)$ can be replaced by $\rho(t)$ which introduces
errors below second order.

We then have
\begin{equation}
{\frac{d\rho}{dt}}=-\int_{0}^{t}d\tau[{{{{{{H_{1}(t),[H_{1}(t-\tau),\rho(t)]}%
}}}}}]\equiv\Gamma\rho(t)
\end{equation}
where $\Gamma$ is the "relaxation matrix", the real parts of which describe
decays of coherence, and the imaginary parts of the off-diagonal elements
describe frequency shifts.

Using the relations in Appendix A together with the expansion of the density
matrix Eq. (\ref{densexpan}), the time-derivative of $\rho$, correct to
second-order and neglecting $2\omega_{0}$ terms, is
\begin{align}
\dot{\rho}_{1,-1}  &  =-\rho_{1,-1}\int_{0}^{t}2e^{i\omega_{0}\tau}b^{\ast
}b^{\prime}d\tau\\
\dot{\rho}_{1,1}  &  =-\rho_{1,1}\int_{0}^{t}2e^{-i\omega_{0}\tau}bb^{\prime
}{}^{\ast}d\tau\\
\dot{\rho}_{1,0}  &  =-\rho_{1,0}\int_{0}^{t}4\mathrm{Re}\left[
e^{i\omega_{0}t}b^{\prime}{}^{\ast}b\right]  d\tau
\end{align}
where
\begin{equation}
2b=\omega_{x}(t)+i\omega_{y}(t);\ \ \ \ 2b^{\prime}=\omega_{x}(t-\tau
)+i\omega_{y}(t-\tau).
\end{equation}
These equations describe both frequency shifts and relaxations of the density
matrix. We are at present most interested in frequency shift, which is given
by the difference in the off-diagonal components of $\Gamma$. Expanding $b$
and $b^{\prime}$ we find
\begin{align}
{\delta\omega(t)}  &  =-{\frac{1}{2}}\int_{0}^{t}\big[  \cos\omega_{0}%
\tau\left(  \omega_{x}(t)\omega_{y}(t-\tau)-\omega_{x}(t-\tau)\omega
_{y}(t)\right) \nonumber\label{devolv}\\
&  +\sin\omega_{0}\tau\left(  \omega_{x}(t)\omega_{x}(t-\tau)+\omega
_{y}(t)\omega_{y}(t-\tau)\right)  \big]  d\tau.
\end{align}
This is the general solution for the frequency shift given an arbitrary
perturbing field. An ensemble average must be taken.

The identical result is obtained with appropriate $\left(  \omega_{x,y}%
,\delta\omega\ll\omega_{o}\right)  $ approximations from the Bloch equation in
the form given in Eqs. (46) and (47) of \cite{JMP}. This is quite interesting
given the different assumptions made in the two approaches.

Now $\omega_{x}=ax+bv_{y},\omega_{y}=ay-bv_{x}$ where%

\begin{align}
a  &  =\frac{\gamma}{2}\frac{\partial B_{z}}{\partial z}\\
b  &  =\gamma\frac{E}{c}%
\end{align}
with $\gamma$ the gyromagnetic ratio and it is clear that only the cross-terms
$\omega_{x}\omega_{y}$ will result in a non-zero linear E $\left(  \varpropto
b\right)  $ shift,%

\begin{align}
\delta\omega &  =-\frac{1}{2}\int_{0}^{t}d\tau\left(  \cos\omega_{o}%
\tau\right)  \left\{  \left\langle \omega_{x}\left(  t\right)  \omega
_{y}\left(  t-\tau\right)  \right\rangle -\left\langle \omega_{x}\left(
t-\tau\right)  \omega_{y}\left(  t\right)  \right\rangle \right\}
\label{skl22}\\
&  =\frac{ab}{2}\int_{0}^{t}d\tau\left(  \cos\omega_{o}\tau\right)  R\left(
\tau\right) \nonumber
\end{align}
where
\begin{equation}
R(\tau)=\Big\langle  y(t)v_{y}(t-\tau)+x(t)v_{x}(t-\tau)-y(t-\tau
)v_{y}(t)-x(t-\tau)v_{x}(t)\Big\rangle  \label{RT}%
\end{equation}
is the net correlation function, where $\langle...\rangle$ represents an
ensemble and time average.

\subsection{General solution for a radial magnetic field plus vxE}

According to (\ref{skl22}) the frequency shift is proportional to the Fourier
transform of the correlation function $R\left(  \tau\right)  ,$ between
$\left(  y,v_{y}\right)  $ and $\left(  x,v_{x}\right)  $ evaluated at the
Larmor frequency, $\omega_{0}$. However this can be written in terms of the
velocity autocorrelation function as follows:%
\begin{align}
y(t)  &  =y_{o}+\int_{0}^{t}v_{y}\left(  t^{\prime}\right)  dt^{\prime
}\nonumber\\
y(t-\tau)  &  =y_{o}+\int_{0}^{t-\tau}v_{y}\left(  t^{\prime}\right)
dt^{\prime}%
\end{align}
Since there are no correlations between $y_{o}$ and $v_{y}$ the $y$ terms in
(\ref{RT}) are%
\begin{align}
A  &  =y(t)v_{y}(t-\tau)=\int_{0}^{t}\left\langle v_{y}\left(  t^{\prime
}\right)  v_{y}(t-\tau)\right\rangle dt^{\prime}\\
B  &  =y(t-\tau)v_{y}(t)=\int_{0}^{t-\tau}\left\langle v_{y}\left(  t^{\prime
}\right)  v_{y}(t)\right\rangle dt^{\prime}\\
R_{y}\left(  \tau\right)   &  =A-B=\nonumber\\
&  =\left(  \int_{0}^{t}\left\langle v_{y}\left(  t^{\prime}\right)
v_{y}(t-\tau)\right\rangle dt^{\prime}-\int_{0}^{t-\tau}\left\langle
v_{y}\left(  t^{\prime}\right)  v_{y}(t)\right\rangle dt^{\prime}\right)
\left(  ab\right) \\
&  =\int_{\tau-t}^{\tau}dx\psi\left(  x\right)  -\int_{\tau}^{t}dx\psi\left(
x\right)
\end{align}
where $\psi\left(  x\right)  $ is the velocity autocorrelation function and we
used the fact that it is an even function of $x$. Repeating the same argument
for the $x$ axis we have
\begin{align*}
\psi\left(  \tau\right)   &  =\left\langle v_{y}\left(  t\right)  v_{y}%
(t-\tau)+v_{x}\left(  t\right)  v_{x}(t-\tau)\right\rangle \\
&  =\left\langle \overrightarrow{v}_{xy}\left(  t\right)  \cdot\overrightarrow
{v}_{xy}(t-\tau)\right\rangle
\end{align*}%
\begin{align}
R\left(  \tau\right)   &  =\int_{\tau-t}^{\tau}dx\psi\left(  x\right)
-\int_{\tau}^{t}dx\psi\left(  x\right) \\
&  =2h\left(  \tau\right)  -h\left(  t-\tau\right)  -h\left(  t\right)
\label{res2a}\\
&  =2h\left(  \tau\right)
\end{align}%
\begin{equation}
h\left(  \tau\right)  =\int_{0}^{\tau}dx\psi\left(  x\right)
\end{equation}
and we consider only cases where $\psi\left(  x\right)  \rightarrow0$ as
$x\rightarrow\infty$ so that we can take the limit $t\rightarrow\infty$ in Eq.
(\ref{res2a}) and we note that a constant term in $R$ will not have any effect
on (\ref{skl22}) contributing only a term $\propto\delta\left(  \omega
_{o}\right)  =0.$

According to (\ref{skl22}) we need the cosine Fourier transform of $R\left(
\tau\right)  $. This will involve $1/\omega$ times the FT of $\psi\left(
x\right)  $ which in turn is proportional to $\omega^{2}$ times the FT of the
position correlation function as we shall see. Substituting (\ref{res2a}) into
(\ref{skl22}) we have

\bigskip%

\begin{equation}
\delta\omega=ab\int_{0}^{t}d\tau\left(  \cos\omega_{o}\tau\right)  h\left(
\tau\right)  \label{ab1}%
\end{equation}

\bigskip$\allowbreak$

Writing the velocity correlation function as
\begin{equation}
\psi(t)=\int_{-\infty}^{\infty}\cos\omega t\psi\left(  \omega\right)
d\omega\label{corr}%
\end{equation}
we have%
\begin{equation}
h\left(  \tau\right)  =\int_{0}^{\tau}\psi(t)dt=\int_{-\infty}^{\infty}%
\psi(\omega)\left(  \frac{\sin\omega\tau-1}{\omega}\right)  d\omega
\label{intpsi}%
\end{equation}
so that according to (\ref{ab1}) the frequency shift is given by (dropping the
time independent term)%
\begin{align}
\delta\omega &  =ab\left[
\begin{array}
[c]{c}%
\int_{0}^{t}d\tau\cos\omega_{o}\tau\int_{-\infty}^{\infty}\psi(\omega
)\frac{\sin\omega\tau}{\omega}d\omega
\end{array}
\right] \nonumber\\
\delta\omega &  =-ab%
\begin{array}
[c]{c}%
\int_{-\infty}^{\infty}\frac{\psi(\omega)}{\left(  \omega_{o}^{2}-\omega
^{2}\right)  }d\omega
\end{array}
\label{res3}%
\end{align}
The equation (\ref{res3}) represents the general solution to our problem which
is simply the single frequency B-S result (Eq. \ref{BS}, \cite{JMP} Eq. (18))
summed over the frequency spectrum of the velocity autocorrelation
function\ plus oscillating terms (omitted) that don't contribute as long as
$\psi\left(  x\right)  \rightarrow0$\ as $x\rightarrow\infty$.

\subsubsection{Example: Particle in circular orbit}

For a particle in an hypothetical circular orbit with orbital frequency
$\omega_{r}\neq\omega_{o}$ we have
\begin{align}
\psi(\tau)  &  =v_{xy}^{2}\cos\omega_{r}\tau\nonumber\\
\psi(\omega)  &  =v_{xy}^{2}\delta\left(  \omega-\omega_{r}\right)
\end{align}
and substituting in (\ref{res3})%
\begin{align}
\delta\omega &  =-ab\int_{-\infty}^{\infty}v_{xy}^{2}\delta\left(
\omega-\omega_{r}\right)  \frac{1}{\omega_{o}^{2}-\omega^{2}}d\omega
+oscillating\text{ }terms\nonumber\\
&  =-\frac{abv_{xy}^{2}}{\omega_{o}^{2}-\omega_{r}^{2}}%
\end{align}
in agreement with (\ref{BS} and Eq. (18) of \cite{JMP}) and valid for the case
when $\left(  \omega_{o}-\omega_{r}\right)  >\omega_{x,y}$.

\section{\bigskip Numerical calculations of the frequency shift}

\subsection{Numerical estimations of the correlation function}

\label{numerical}

The problem of the neutron EDM experiment with a $^{199}$Hg comagnetometer
subject to a time-varying $v\times E$ field in combination with a
spatially-varying magnetic field is described in \cite{JMP} and in the
Introduction. We assume a cylindrical volume with radial field
$\overrightarrow{B}(r)=a^{\prime}r\hat{r}$. The electric field is constant
everywhere and along the $\hat{z}$ direction. Assuming a constant velocity $v
$, the $v\times E$ field is then fluctuating in direction but of spatially
uniform magnitude.

A numerical calculation of the correlation function was performed for the
two-dimensional case (UCN or Hg at a fixed $z$, moving only in $x-y$ plane).
This problem can be parameterized in terms of the time between collisions
$\tau_{c}=\lambda/v$, where the mean free path between collisions is $\lambda$
and the average velocity is $v$. For the numerical calculations, $v$ is
assumed constant. Time can be parameterized in dimensionless units, $\tau
/\tau_{c}$. The correlation function was calculated by statistically choosing
a propagation distance for a fixed velocity direction, and taking time steps
of \ $0.025$, after which a new random velocity direction was chosen. Various
degrees of specularity, parameterized by $\Delta\theta$ for the statistical
degree of angular change for reflection from the bottle surface, were considered.

Results of a two-dimensional Monte Carlo calculation are shown in Figure 1.
Taking $\lambda=1$ and fixed, we see the effect of wall collisions as the
bottle radius approaches $\lambda$. We see in Fig. 1 that in all cases
$R(\tau)$ initially increases linearly. The effect of the wall collisions when
$R>\lambda$ is to limit the distance that the random walk can take, and this
appears as an exponential decay in $R\left(  \tau\right)  $ at long times.
This effect does not depend on the specularity of the wall collisions and is
best seen as an effect on the whole ensemble of particles which can be
described by classical diffusion theory. In this limit, the correlation
function is well-described by
\begin{equation}
R(\tau)=(1-e^{-\tau/\tau_{c}})e^{-\tau/T} \label{bigg}%
\end{equation}
where, from analysis of the plots,
\begin{equation}
T\approx{\frac{0.6R^{2}}{\lambda v}}. \label{tdiff}%
\end{equation}

In the other limit, $R<\lambda$, $R(\tau)$ oscillates with frequency
\begin{equation}
\omega\approx{\frac{2\pi v}{5.2R}} \label{omega}%
\end{equation}
and
\begin{equation}
R(\tau)=e^{-\tau/T}\sin\omega\tau\label{small}%
\end{equation}
where $T$ depends on $\Delta\theta$, but is typically of order $2\pi/\omega$.

The frequency shift is determined by Eq. (\ref{skl22}) and in the case of
large $R$ we find ($\tau_{c}<<T$), using Eq. (\ref{tdiff})
\begin{equation}
\delta\omega={\frac{abR^{2}}{T^{2}\omega_{0}^{2}+1}}={\frac{abR^{2}%
}{1+(0.6R^{2}\omega_{0}/v\lambda)^{2}}}.
\end{equation}
These results are in good agreement with \cite{JMP}, Fig. 10, for which
$4/2\pi\approx0.634$ replaces the factor $0.6$ above and with Eq. (\ref{oond}) below.

Additional insight can be gained by considering the effects of varying
$\lambda$ keeping $R$ fixed, as shown in Fig. 2 for very small $\lambda$. In
this limit, the horizontal axis is multiplied by $\lambda/R$ to define time
proportional to $R/v$. The correlation amplitude function is proportional to
$\lambda v$ and the decaying exponential time constant is
\begin{equation}
T\propto{\frac{R^{2}}{\lambda v}}.
\end{equation}
The time to reach the peak value is
\begin{equation}
\tau_{0}\propto\lambda^{2}/Rv
\end{equation}
which approaches zero as $R\rightarrow\infty$.

This limit is further discussed in Sec. 4.1, and the frequency shift in this
case is in general agreement with Fig. 10 of \cite{JMP}.

The curves for large (relative to $R$) $\lambda$ in figure 1 show damped
oscillations whose damping depends on the angular spread of the wall
collisions. This is a manifestation of the resonance behavior discussed in
ref. \cite{JMP} for the case of perfectly specular wall collisions. Here we
see the damping due to non-specular reflections.

\subsection{Numerical estimations of the frequency shift for all values of
$\omega_{o}/\omega_{r}$}

Using Eq. (\ref{ab1}), and the results of the previous section, the cosine
transform of the numerically-determined correlation function can be calculated
numerically. In order to reduce oscillations due to the finite time window, a
Hamming window function was applied to the correlation function, and a slight
correction due to the frequency dependent gain as imposed by the window
function was applied. The results, as a function of mean free path $\lambda$
at fixed radius $R$, for specular and purely diffuse wall reflection, are
shown in Fig. 3.

There are a few points worth noting. First, the curves for large $\lambda$ in
the specular case are very similar to the Bloch-Siegert result. Second, at
small and large frequencies, the results agree with the numerical
semi-analytically determined results presented above, and in \cite{JMP} and
the theoretical analysis below. Third, the behavior at intermediate
frequencies is seen to be very interesting: The shift goes to zero for
$\omega_{o}/\omega_{r}\sim1$ as it must because the effect changes sign
between large and small frequencies.

Furthermore, it can be seen immediately that the effects of wall collision
specularity is important when $\omega_{o}\approx\omega_{r}$, in contradiction
to the statement in \cite{JMP} that the degree of specularity does not affect
the frequency shift. We discuss this point later in more detail (Sec. IV).

\section{Analytical results for the limiting cases of large and small
frequencies $(\omega_{o}/\omega_{r}\gg1,\omega_{o}/\omega_{r}\ll1)$}

Equation (\ref{res3} ) represents the formal solution of the problem in all
cases of interest here. Thus the frequency shift is determined entirely by the
velocity auto-correlation function of the particles undergoing magnetic
resonance. This function has been the subject of intense experimental and
theoretical study (\cite{Egelstaff, Lovesey,squires}). In our case, involving
macroscopic distances and times, it suffices to treat the motion classically.
For relatively short times if the particles undergo collisions which are
distributed according to a Poisson distribution with average time between
collisions given by $\tau_{c}$, the velocity correlation function is well
known to be given by%
\begin{equation}
\psi\left(  t\right)  =\left\langle v^{2}\right\rangle e^{-t/\tau_{c}}
\label{exp}%
\end{equation}

This form is known to be valid for relatively short times. According to Eq.
(\ref{ab1}) the frequency shift depends on the Fourier transform of the
integral of the velocity correlation function evaluated at $\omega_{o}$. So
the short time behavior of $\psi\left(  t\right)  $ determines the high
frequency behavior of $\psi\left(  \omega\right)  $, and the result using this
form is expected be valid in the case of large $\omega_{o}$, i.e. $\omega
_{o}\gg\omega_{r}$.

For longer times the velocity correlation function is well described by
classical diffusion theory. Thus the long time behavior will determine the low
frequency region of the velocity spectrum and the result will apply to the
case $\omega_{o}\ll\omega_{r}$. In this region the result will depend on the
size of the containing vessel as the dynamics of the diffusion process are
influenced by the boundary conditions.

\subsection{Short correlation times ($\omega_{r}\ll\omega_{o} $)}

Using (\ref{exp}) we have%
\begin{equation}
\psi\left(  \omega\right)  =\frac{1}{\pi}\left\langle v^{2}\right\rangle
\int_{0}^{\infty}\cos\omega te^{-t/\tau}dt=\frac{1}{\pi}\left\langle
v^{2}\right\rangle \frac{1}{\tau\left(  \omega^{2}+\frac{1}{\tau^{2}}\right)
}%
\end{equation}
so that according to (\ref{res3})
\begin{align}
\delta\omega &  =-ab\int_{-\infty}^{\infty}\frac{\psi(\omega)}{\left(
\omega_{o}^{2}-\omega^{2}\right)  }d\omega\\
&  =ab\frac{1}{\pi}\frac{\left\langle v^{2}\right\rangle }{\tau}\int_{-\infty
}^{\infty}\frac{1}{\left(  \omega^{2}+\frac{1}{\tau^{2}}\right)  \left(
\omega^{2}-\omega_{o}^{2}\right)  }d\omega\\
&  =-ab\frac{\left\langle v^{2}\right\rangle }{\omega_{o}^{2}}\frac{1}{\left(
1+\frac{1}{\omega_{o}^{2}\tau^{2}}\right)  } \label{aaa}%
\end{align}
This is in substantial agreement with \ the expression given in the caption of
\cite{JMP} Fig. 12 when it is taken with \cite{JMP} Eq. (19) or (\ref{2})
applicable to the case when $\omega_{r}\ll\omega_{o}$. It is quite likely that
the small discrepancy ($\sim10\%$) in the 50\% suppression point is due to the
process of averaging over the velocity distribution in \cite{JMP} Fig. 12.

$\allowbreak$

\subsection{\label{theorem}Diffusion theory calculation of the long time
behavior of the velocity correlation function. Frequency shifts for
($\omega_{r}\gg\omega_{o}$)}

Whereas the previous case applies to UCN this case would apply to atoms used
as a comagnetometer and is more relevant experimentally as it results in
larger shifts \cite{JMP} and in some cases \cite{RPPedm} the collision rate
can be simply adjusted by changing the experimental conditions.

In the following we review the solution of the diffusion equation in
cylindrical geometry, obtain the velocity autocorrelation function from the
solution and calculate the frequency shift. In the limit of small collision
rate the result agrees with the known results for ($\omega_{r}\gg\omega_{o}$)
and the effect of the collisions agrees with that found from numerical
simulations (\cite{JMP} Fig. 10)

\subsubsection{Green's function for the diffusion equation in cylindrical geometry}

In this section we attempt to understand the effects of the vessel boundary on
the velocity autocorrelation function, observed in the numerical simulations
(section \ref{numerical}), by \ applying classical diffusion theory to the
problem. Diffusion theory is expected to be valid for long times so that we
expect the results to be valid for small $\omega_{o}$, i.e. $\omega_{o}%
\ll\omega_{r}$.
\begin{align}
D\nabla^{2}\rho-\frac{\partial\rho}{\partial t}  &  =0\nonumber\\
\rho &  =u_{k}(r)e^{-Dk^{2}t}\nonumber\\
\nabla^{2}u+k^{2}u  &  =0
\end{align}

We consider a two dimensional problem, that is we neglect any $z$ dependence
$\left(  k_{z}=0\right)  .$ For the cases considered in \cite{JMP} where the
height of the bottle is much smaller than the radius higher $z$ modes will
decay relatively quickly.

The boundary condition is $j\left(  R\right)  =-D\frac{\partial\rho}{\partial
r}=0$ so the eigenfunctions satisfying the boundary conditions are%
\begin{align*}
u_{m,n}  &  =N_{m,n}J_{m}(k_{m,n}r)e^{im\theta}\\
k_{m,n}R  &  =x_{m,n}^{\prime}\text{ (}n^{th}\text{ zero of }dJ_{m}%
(z)/dz\text{)}%
\end{align*}
where the normalization constant (which depends on the boundary conditions) is
(\cite{Sommerfeld}, p 322)%
\begin{equation}
N_{m,n}=\frac{1}{\sqrt{2\pi}J_{m}(k_{m,n}R)}\sqrt{\frac{2k_{m,n}^{2}}{\left(
k_{m,n}R\right)  ^{2}-m^{2}}} \label{norm}%
\end{equation}

\bigskip The Greens' function satisfying the boundary conditions is
\cite{morsefesh}%
\begin{equation}
G(r,r^{\prime},t)=\sum_{m,n}\left(  N_{m,n}\right)  ^{2}J_{m}(k_{m,n}%
r)J_{m}(k_{m,n}r^{\prime})e^{im\left(  \theta-\theta^{\prime}\right)
}e^{-Dk_{m,n}^{2}t} \label{green}%
\end{equation}
This is the probability of finding a particle at $\overrightarrow{r}$ at time
$t$, given that the particle was at $\overrightarrow{r}^{\prime}$ at time
$t=0$. The spectrum of the velocity correlation function is related to
$S\left(  \overrightarrow{q},\omega\right)  $ which in turn is the average
over the system of the Fourier transform of this probability with respect to
$\rho=\left(  r-r^{\prime}\right)  $. We use the cosine transform because we
want the cosine transform of the velocity correlation function (\ref{corr}%
)\emph{\ }$\psi\left(  \omega\right)  $.%

\begin{align}
S(q,\omega)  &  =\frac{1}{\pi}\left\langle \int d^{2}\rho e^{i\overrightarrow
{q}\cdot\overrightarrow{\rho}}\int_{0}^{\infty}dt\cos\omega tG(r,r^{\prime
},t)\right\rangle \\
&  =\frac{1}{\pi}\int\int\frac{d^{2}r^{\prime}}{\pi R^{2}}d^{2}\rho
e^{i\overrightarrow{q}\cdot\overrightarrow{\rho}}\int_{0}^{\infty}dt\cos\omega
tG(r,r^{\prime},t)\\
&  =\frac{1}{\pi^{2}R^{2}}\sum_{m,n}\left(  N_{m,n}\right)  ^{2}\int
d^{2}re^{i\overrightarrow{q}\cdot\overrightarrow{r}}J_{m}(k_{m,n}%
r)e^{im\theta}\times\\
&  \int d^{2}r^{\prime}e^{-i\overrightarrow{q}\cdot\overrightarrow{r}^{\prime
}}J_{m}(k_{m,n}r^{\prime})e^{-im\theta^{\prime}}\frac{Dk_{m,n}^{2}}{\omega
^{2}+\left(  Dk_{m,n}^{2}\right)  ^{2}}%
\end{align}
Now we can evaluate the integrals using
\begin{equation}
J_{m}\left(  x\right)  =\frac{\left(  -i\right)  ^{m}}{2\pi}\int_{0}^{2\pi
}e^{i\left(  x\cos\theta+m\theta\right)  }d\theta
\end{equation}
and Bessel function identities
\begin{align*}
\int d^{2}re^{i\overrightarrow{q}\cdot\overrightarrow{r}}J_{m}(k_{m,n}%
r)e^{im\theta}  &  =\frac{2\pi\left(  i\right)  ^{\pm m}}{\left(
q^{2}-k_{m,n}^{2}\right)  }\frac{qR}{2}J_{m}(k_{m,n}R)\times\\
&  \left[  J_{m-1}(qR)-J_{m+1}(qR)\right]
\end{align*}

thus%
\begin{align}
S(q,\omega)  &  =\frac{2}{\pi^{3}R^{2}}\sum_{m,n}..\\
&  \frac{k_{m,n}^{2}}{\left(  \left(  k_{m,n}R\right)  ^{2}-m^{2}\right)
}\left(  \frac{2\pi}{\left(  q^{2}-k_{m,n}^{2}\right)  }\frac{qR}{2}\left[
J_{m-1}(qR)-J_{m+1}(qR)\right]  \right)  ^{2}\frac{Dk_{m,n}^{2}}{\left(
\omega^{2}+\left(  Dk_{m,n}^{2}\right)  ^{2}\right)  } \label{4b}%
\end{align}

\subsubsection{Velocity autocorrelation function}

The velocity autocorrelation function
\begin{equation}
\psi(\tau)=\left\langle \overrightarrow{v}\left(  0\right)  \cdot
\overrightarrow{v}\left(  \tau\right)  \right\rangle
\end{equation}
has a Fourier transform given by (\cite{degenn})%
\begin{equation}
\psi\left(  \omega\right)  =\lim_{q\rightarrow0}2\left(  \frac{\omega}%
{q}\right)  ^{2}S(q,\omega) \label{THeorem}%
\end{equation}
so that the only terms in (\ref{4b}) which contribute are those containing
$J_{0}\left(  qR\right)  $, since $\lim_{x\rightarrow0}J_{n}\left(  x\right)
\thicksim\left(  x\right)  ^{n}$, $J_{0}\left(  0\right)  =1.$ Thus we only
need to keep terms with $m=\pm1$ in (\ref{4b}) and we find%

\begin{equation}
\lim_{q\rightarrow0}S(q,\omega)=\frac{2q^{2}}{\pi\left(  \left(
k_{1,n}R\right)  ^{2}-1\right)  }\frac{D}{\left(  \omega^{2}+\left(
Dk_{1,n}^{2}\right)  ^{2}\right)  }%
\end{equation}

\bigskip\ Then%
\begin{equation}
\psi\left(  \omega\right)  =\frac{1}{\pi}\sum_{n}\frac{4}{\left(  x_{1,n}%
^{2}-1\right)  }\frac{D\omega^{2}}{\left(  \omega^{2}+\left(  Dk_{1,n}%
^{2}\right)  ^{2}\right)  } \label{psi(W)}%
\end{equation}

\subsubsection{Frequency shift in the diffusion approximation (cylindrical geometry)}

According to (\ref{res3})%

\begin{align}
\delta\omega &  =-ab\int_{-\infty}^{\infty}\psi(\omega)\frac{1}{\omega_{o}%
^{2}-\omega^{2}}d\omega\\
&  =abR^{2}\sum_{n}\frac{4}{\left(  x_{1,n}^{2}-1\right)  }\frac{1}%
{x_{1,n}^{2}\left(  \left(  \frac{\omega_{o}R^{2}}{Dx_{1,n}^{2}}\right)
^{2}+1\right)  } \label{big}%
\end{align}

The result (\ref{big}) is dominated by the first mode $x_{1,1}=1.84$. Figure 3
shows the first term in comparison to the sum of the first 4 terms. For
convenience we list the zeroes of $J_{1}^{\prime}(x)$: $x_{1,2}=5.33,x_{1,3}%
=8.54,x_{1,4}=11.7$. Since we are dealing with a 2 dimensional problem we put%
\begin{equation}
D=v^{2}\tau/2
\end{equation}
(instead of $\tau v^{2}/3$ for 3 dimensions) in order to facilitate the
comparison with the numerical simulations \cite{JMP} and obtain for the
condition that the frequency shift is reduced to 50\% of its value in the
absence of collisions%
\begin{equation}
\eta=\frac{\omega_{o}R^{2}}{Dx_{1,n}^{2}}=\frac{2\omega_{o}R^{2}}{v^{2}\tau
x_{1,n}^{2}}=.59\frac{\omega_{o}R^{2}}{v^{2}\tau}=1 \label{oond}%
\end{equation}
the numerical factor of which is to be compared with $\frac{2}{\pi}=.634$
obtained in [\cite{JMP}], fig. 10 by fitting simulated results, and our
numerical result of $0.6$ presented in Sec. \ref{numerical}. The magnitude
$abR^{2}/2$ of (\ref{big}) in the absence of collisions is just that expected
from the Bloch-Siegert treatment in the case $\omega_{r}\gg\omega_{o}$ (Eq.
(\ref{2}), \cite{JMP} Eq. (21))$,$ averaged over the different trajectories as
discussed in \cite{JMP} after equation (22).

\subsubsection{Frequency shift in the diffusion approximation (rectangular geometry)}

For the rectangular case the normalized eigenfunctions are%
\begin{equation}
u_{m,n}\left(  x,y\right)  =\sqrt{\frac{2}{L_{x}}}\cos\frac{m\pi}{L_{x}}%
x\sqrt{\frac{2}{L_{y}}}\cos\frac{m\pi}{L_{y}}y
\end{equation}
which satisfy the reflection boundary conditions at $x=0,L_{x}$ and
$y=0,L_{y}$. For $n$ or $m=0$ the corresponding eigenfunctions are
\begin{equation}
u_{0}=\frac{1}{\sqrt{L_{x,y}}}%
\end{equation}

so that the Green's function is
\begin{align}
G(x,x^{\prime},y,y^{\prime},t)  &  =\sum_{m,n=0}^{\infty}\left[
\frac{1}{L_{x}}+\frac{2}{L_{x}}\cos k_{m}x\cos k_{m}x^{\prime}e^{-Dk_{m}^{2}%
t}\right]  \times\nonumber\\
&  \left[  \frac{1}{L_{y}}+\frac{2}{L_{y}}\cos k_{n}x\cos k_{n}x^{\prime
}e^{-Dk_{n}^{2}t}\right]
\end{align}
with $k_{m,n}=\left(  m,n\right)  \pi/L_{x,y}$. To calculate $\lim
_{q\rightarrow0}S\left(  q,\omega\right)  $ we need integrals of the form
\begin{align}
\lim_{q\rightarrow0}\int_{0}^{L}e^{iq_{x}x}\cos k_{m}xdx  &  =\frac{q_{x}%
}{q_{x}^{2}-k_{m}^{2}}\left\{
\begin{array}
[c]{cc}%
q_{x}L_{x} & m=2,4,6..\\
\frac{-2}{i} & m=1,3,5..
\end{array}
\right\} \label{qqq}\\
&  =\left\{
\begin{array}
[c]{cc}%
L_{x} & m=0
\end{array}
\right\}
\end{align}
Since each of these will appear squared because of the contribution from the
$x,x^{\prime}$ integrals we can only take the odd values of $m$. The even
numbers will yield a higher power of $q$ which will vanish in the limit. Given
this, if we take $m=1,3,5$ we must take $n=0$ and vice-versa. We calculate,
using (\ref{qqq})%
\begin{align}
&  \lim_{q\rightarrow0}\left[  S\left(  q,\omega\right)  =\int dx\int
\frac{dx^{\prime}}{L_{x}}\int dy\int\frac{dy^{\prime}}{L_{y}}%
e^{i\overrightarrow{q}\cdot\left(  \overrightarrow{x}-\overrightarrow
{x}^{\prime}\right)  }\frac{1}{\pi}\int_{0}^{\infty}dt\cos\omega tG\left(
\overrightarrow{x},\overrightarrow{x}^{\prime},t\right)  \right] \\
&  =q^{2}\frac{8}{2\pi}\left(
\begin{array}
[c]{c}%
\sum_{m=1,3,5..}\frac{1}{k_{m}^{2}\left(  m\pi\right)  ^{2}}\frac{Dk_{m}^{2}%
}{\omega^{2}+\left(  Dk_{m}^{2}\right)  ^{2}}+\\
\sum_{n=1,3,5..}\frac{1}{k_{n}^{2}\left(  n\pi\right)  ^{2}}\frac{Dk_{n}^{2}%
}{\omega^{2}+\left(  Dk_{n}^{2}\right)  ^{2}}%
\end{array}
\right)
\end{align}
where we used $\left\langle q_{x}^{2}\right\rangle =\left\langle q_{y}%
^{2}\right\rangle =q^{2}/2$. Then%
\begin{equation}
\psi\left(  \omega\right)  =\frac{8\omega^{2}}{\pi}\left(  \sum_{m=1,3,5..}%
\frac{1}{k_{m}^{2}\left(  m\pi\right)  ^{2}}\frac{Dk_{m}^{2}}{\omega
^{2}+\left(  Dk_{m}^{2}\right)  ^{2}}+\sum_{n=1,3,5..}\frac{1}{k_{n}%
^{2}\left(  n\pi\right)  ^{2}}\frac{Dk_{n}^{2}}{\omega^{2}+\left(  Dk_{n}%
^{2}\right)  ^{2}}\right)
\end{equation}
and (using \ref{res3})%
\begin{align}
\delta\omega &  =-ab\int_{-\infty}^{\infty}\psi(\omega)\frac{1}{\omega_{o}%
^{2}-\omega^{2}}d\omega\\
&  =8ab\left(  \sum_{m=1,3,5..}\frac{L_{x}^{2}}{\left(  m\pi\right)  ^{4}%
}\frac{1}{\left(  \frac{\omega_{o}L_{x}^{2}}{D\left(  m\pi\right)  ^{2}%
}\right)  ^{2}+1}+\sum_{n=1,3,5..}\frac{L_{y}^{2}}{\left(  n\pi\right)  ^{4}%
}\frac{1}{\left(  \frac{\omega_{o}L_{y}^{2}}{D\left(  n\pi\right)  ^{2}%
}\right)  ^{2}+1}\right)
\end{align}
We thus see that in a rectangular box $L_{x}\neq L_{y}$ it is the longer side
which dominates the behavior.

\subsection{\label{notheorem}Diffusion theory calculation of the long time
behavior of the velocity correlation function. Frequency shifts for
($\omega_{r}\gg\omega_{o}$) (Alternate caclulation)}

In this section we derive the diffusion theory result for cylindrical geometry
(\ref{big}) using an alternate method based on that of Mcgregor
\cite{McGregor} which avoids the use of the theorem (\ref{THeorem}). We start
with the Green's function for cylindrical geometry given above (\ref{green})%

\begin{equation}
G(r,r^{\prime},t)=\sum_{m,n}\left(  N_{m,n}\right)  ^{2}J_{m}(k_{m,n}%
r)J_{m}(k_{m,n}r^{\prime})e^{im\left(  \phi-\phi^{\prime}\right)
}e^{-Dk_{m,n}^{2}t}%
\end{equation}
This is the probability of finding a particle at $\overrightarrow{r}$ at time
$t$, given that the particle was at $\overrightarrow{r}^{\prime}$ at time
$t=0$.

\subsubsection{Position-velocity correlation function $R(\tau)$}

From equ. (\ref{RT}) we have%
\begin{align}
R(\tau)  &  =\left\langle y(t)v_{y}(t-\tau)+x(t)v_{x}(t-\tau)-y(t-\tau
)v_{y}(t)-x(t-\tau)v_{x}(t)\right\rangle \\
&  =\left\langle \overrightarrow{r}(t)\cdot\overrightarrow{v}(t-\tau
)-\overrightarrow{r}(t-\tau)\cdot\overrightarrow{v}(t\right\rangle =\left|
\frac{d}{dt^{\prime}}\left\langle \overrightarrow{r}(t)\cdot\overrightarrow
{r}(t\prime)\right\rangle -\frac{d}{dt}\left\langle \overrightarrow{r}%
(t\prime)\cdot\overrightarrow{r}(t)\right\rangle \right|  _{t^{\prime}=t-\tau}
\label{R=dr/dt}%
\end{align}

Following McGregor \cite{McGregor} we write
\begin{align}
\left\langle \overrightarrow{r}(t)\cdot\overrightarrow{r}(t\prime
)\right\rangle  &  =\int d^{2}r\int\frac{d^{2}r^{\prime}}{\pi R^{2}%
}G(r,r^{\prime},t-t^{\prime})\left\langle \overrightarrow{r}\cdot
\overrightarrow{r^{\prime}}\right\rangle \\
&  =\frac{1}{\pi R^{2}}\sum_{m,n}\left(  N_{m,n}\right)  ^{2}\int rdrd\phi
J_{m}(k_{m,n}r)\int r^{\prime}dr^{\prime}d\phi^{\prime}J_{m}(k_{m,n}r^{\prime
})\times\\
&  \times rr^{\prime}e^{im\left(  \phi-\phi^{\prime}\right)  }\left[  \cos
\phi\cos\phi^{\prime}+\sin\phi\sin\phi^{\prime}\right]  e^{-Dk_{m,n}%
^{2}\left(  t-t^{\prime}\right)  }%
\end{align}
Since $\left(  N_{m,n}\right)  ^{2}=\left(  N_{-m,n}\right)  ^{2}$ and
$J_{-m}(z)=(-1)^{m}J_{m}(z),$ (\ref{norm}) we can (excluding $m=0$, which will
be seen not contribute to the result) combine the terms for $m$ and $-m$ as follows:%

\begin{align}
\left\langle \overrightarrow{r}(t)\cdot\overrightarrow{r}(t\prime
)\right\rangle  &  =\frac{1}{\pi R^{2}}\sum_{m>0,n}\left(  N_{m,n}\right)
^{2}\int r^{2}drJ_{m}(k_{m,n}r)\int r^{\prime^{2}}dr^{\prime}J_{m}%
(k_{m,n}r^{\prime})\times\\
&  \times\int d\phi\int d\phi^{\prime}\Xi(\phi,\phi^{\prime})e^{-Dk_{m,n}%
^{2}\left(  t-t^{\prime}\right)  }%
\end{align}
where%
\begin{align}
\Xi(\phi,\phi^{\prime})  &  =\left[  \cos\phi\cos\phi^{\prime}+\sin\phi
\sin\phi^{\prime}\right]  2\cos m\left(  \phi-\phi^{\prime}\right) \\
&  =2\left[  \cos\phi\cos\phi^{\prime}+\sin\phi\sin\phi^{\prime}\right]
\left[  \cos m\phi\cos m\phi^{\prime}+\sin m\phi\sin m\phi^{\prime}\right] \\
&  =\frac{1}{2}\left(  \cos\phi\cos m\phi\cos\phi^{\prime}\cos m\phi^{\prime
}+\sin\phi\sin m\phi\sin\phi^{\prime}\sin m\phi^{\prime}+\right. \\
&  \left.  +\sin\phi\cos m\phi\sin\phi^{\prime}\cos m\phi^{\prime}+\cos
\phi\sin m\phi\cos\phi^{\prime}\sin m\phi^{\prime}\right) \\
&  =\frac{1}{2}\left[  \left(  \cos\left(  m-1\right)  \phi+\cos\left(
m+1\right)  \phi\right)  \left(  \cos\left(  m-1\right)  \phi^{\prime}%
+\cos\left(  m+1\right)  \phi^{\prime}\right)  +\right. \\
&  +\left(  \cos\left(  m-1\right)  \phi-\cos\left(  m+1\right)  \phi\right)
\left(  \cos\left(  m-1\right)  \phi^{\prime}-\cos\left(  m+1\right)
\phi^{\prime}\right)  +\\
&  +\left(  \sin\left(  m-1\right)  \phi-\sin\left(  m+1\right)  \phi\right)
\left(  \sin\left(  m-1\right)  \phi^{\prime}-\sin\left(  m+1\right)
\phi^{\prime}\right)  +\\
&  +\left(  \sin\left(  m-1\right)  \phi+\sin\left(  m+1\right)  \phi\right)
\left(  \sin\left(  m-1\right)  \phi^{\prime}+\sin\left(  m+1\right)
\phi^{\prime}\right)
\end{align}
using
\begin{align}
\int_{0}^{2\pi}d\phi\cos(m+1)\phi &  =0\quad(m>0),\quad\int_{0}^{2\pi}%
d\phi\sin(m\pm1)\phi=0\\
\int_{0}^{2\pi}d\phi\cos(m-1)\phi &  =2\pi\delta_{m1}%
\end{align}
we have%
\begin{equation}
\int d\phi\int d\phi^{\prime}\Xi(\phi,\phi^{\prime})=\frac{\left(
2\pi\right)  ^{2}}{2}2\delta_{m1}%
\end{equation}
so that%
\begin{align}
\left\langle \overrightarrow{r}(t)\cdot\overrightarrow{r}(t\prime
)\right\rangle  &  =\frac{1}{\pi R^{2}}\left(  2\pi\right)  ^{2}\sum
_{m>0,n}\delta_{m1}\left(  N_{m,n}\right)  ^{2}\int r^{2}drJ_{m}(k_{m,n}r)\int
r^{\prime^{2}}dr^{\prime}J_{m}(k_{m,n}r^{\prime})e^{-Dk_{m,n}^{2}\left(
t-t^{\prime}\right)  }\\
&  =\frac{4\pi}{R^{2}}\sum_{n}\left(  N_{1,n}\right)  ^{2}\int_{0}^{R}%
r^{2}drJ_{1}(k_{1,n}r)\int_{0}^{R}r^{\prime^{2}}dr^{\prime}J_{1}%
(k_{1,n}r^{\prime})e^{-Dk_{1,n}^{2}\left(  t-t^{\prime}\right)  }%
\end{align}
Using%
\begin{align}
\int_{0}^{\alpha}z^{n}dzJ_{n-1}(z)  &  =\left[  z^{n}J_{n}(z)\right]
_{0}^{\alpha}\\
J_{n}^{\prime}(z)  &  =\frac{n}{z}J_{n}(z)-J_{n+1}(z)
\end{align}
we have%
\begin{equation}
\int_{0}^{R}r^{2}drJ_{1}(k_{1,n}r)=\frac{R^{3}}{\left(  x_{1,n}^{\prime
}\right)  ^{2}}J_{1}(x_{1,n}^{\prime})
\end{equation}
Thus%
\begin{align}
\left\langle \overrightarrow{r}(t)\cdot\overrightarrow{r}(t\prime
)\right\rangle  &  =\frac{4\pi}{R^{2}}\sum_{n}\left(  N_{1,n}\right)
^{2}\frac{R^{6}}{\left(  x_{1,n}^{\prime}\right)  ^{4}}J_{1}^{2}%
(x_{1,n}^{\prime})e^{-Dk_{1,n}^{2}\left(  t-t^{\prime}\right)  }\\
&  =4R^{2}\sum_{n}\frac{1}{\left(  x_{1,n}^{\prime}\right)  ^{2}\left(
\left(  x_{1,n}^{\prime}\right)  ^{2}-1\right)  }e^{-Dk_{1,n}^{2}\left(
t-t^{\prime}\right)  }\\
&  =\sum_{n}\eta_{n}e^{-Dk_{1,n}^{2}\left(  t-t^{\prime}\right)  }%
\end{align}
and%
\begin{equation}
R\left(  \tau\right)  =2\sum_{n}\eta_{n}Dk_{1,n}^{2}e^{-Dk_{1,n}^{2}\tau}%
\end{equation}
using (\ref{R=dr/dt}). Then according to (\ref{skl22})%
\begin{align}
\delta\omega &  =\frac{ab}{2}\int_{0}^{t}d\tau\left(  \cos\omega_{o}%
\tau\right)  R\left(  \tau\right) \\
&  =ab\int_{0}^{t}d\tau\left(  \cos\omega_{o}\tau\right)  \sum_{n}\eta
_{n}Dk_{1,n}^{2}e^{-Dk_{1,n}^{2}\tau}\\
&  =4abR^{2}\sum_{n}\frac{1}{\left(  x_{1,n}^{\prime}\right)  ^{2}\left[
\left(  x_{1,n}^{\prime}\right)  ^{2}-1\right]  }\frac{1}{\left[  \left(
\frac{\omega_{o}R^{2}}{Dx_{1,n}^{\prime2}}\right)  ^{2}+1\right]  }
\label{biggie}%
\end{align}
in complete agreement with equation (\ref{big}).

\subsection{\bigskip Application: $^{3}$He Comagnetometer}

In \cite{RPPedm} the use of $^{3}$He as a comagnetometer for a UCN neutron EDM
experiment is discussed. This system is rather unique in that an effective
background gas (phonons) can be introduced which affects the $^{3}$He
significantly while having no substantial interaction with the UCN for
temperatures below 0.5 K. Because the $^{3}$He and neutron magnetic moments
are equal to within 10\%, it is possible to control this systematic by varying
the size of the effect for $^{3}$He by changing the diffusion rate of the
$^{3}$He.

The UCN upscattering lifetime varies as $100T^{-7}$ s for $T<0.7$ K, while the
coefficient of diffusion for $^{3}$He in a superfluid helium bath varies as
$D\approx1.6T^{-7}$ cm$^{2}$/s \cite{diffusion}.

\bigskip In connection with (\ref{oond}) this yields $\eta=1$ when the
superfluid helium temperature is $T\approx0.25$ K, (R=25cm), which determines
the temperature scale where the effect can be varied, and is within the design
range of operating temperature for the planned experiment, compatible with a
UCN upscattering lifetime in excess of 1000 s.

\section{Discussion}

One of the surprising, but unexplained results of \cite{JMP} was that
according to their numerical simulations, wall collisions had no influence on
the magnitude of the frequency shifts while gas collisions could eliminate the
frequency shifts completely if their rate is high enough. This was apparently
only studied in the limits of large and small $\omega_{o}/\omega_{r}.$ We now
know that this does not apply to intermediate frequencies, e.g, when
$\omega_{o}\sim\omega_{r}$. In Fig. 3 we see that wall collisions have a
serious influence at intermediate frequencies when $\lambda\geq R$. Also from
Fig. 3 we see that the curves for diffuse wall reflections in the absence of
gas collisions is very similar to the specular curves for $\lambda< R/2$. This
implies that there is no essential difference between wall and gas collisions.
We now show that the reason the wall collisions have no effect at the limiting
frequencies, contrary to the case at intermediate frequencies, is that the
wall collisions are never fast enough to influence the systematic
(proportional to $\overrightarrow{E})$ frequency shifts in the limits of large
and small $\omega_{o}$.

For a particles in a cylindrical vessel following a trajectory along a chord
subtending an angle $2\alpha$, the time between collisions is%
\begin{equation}
\tau_{c}=\frac{2R}{v}\sin\alpha
\end{equation}
and the effective field rotation frequency is given by
\begin{equation}
\omega_{r}=2\alpha/\tau_{c}=\frac{\alpha v}{R\sin\alpha}%
\end{equation}
Considering first the case when $\omega_{r}\gg\omega_{o}$ (\ref{big}%
,\cite{JMP} Fig. 10,) the systematic frequency shift was found to be
suppressed by the factor $\eta$%
\begin{align*}
\eta &  =\frac{1}{1+\beta^{2}}\\
\beta &  =\frac{2R^{2}\omega_{o}}{\pi v^{2}\tau_{c}}%
\end{align*}
For significant suppression we need $\beta\gtrsim1$%
\begin{align}
\frac{2R^{2}\omega_{o}}{\pi v^{2}\tau_{c}}  &  =\frac{R\omega_{o}}{\pi
v\sin\alpha}\gtrsim1\\
\sin\alpha &  \lesssim\frac{R\omega_{o}}{\pi v}=\frac{25\times2\cdot7}{10^{4}%
}\sim\frac{1}{30}\nonumber
\end{align}
for representative conditions in \cite{JMP}, fig. 10. ($R=25,B_{o}=1\mu
T,v=10^{4}cm/\sec$).

The probability of a given value of $\alpha$ is given in Eq. (B1) of
\cite{JMP} as
\begin{align}
P(\alpha)d\alpha &  =\frac{4}{\pi}\sin^{2}\alpha d\alpha\\
P(\alpha &  \leq\varepsilon)\sim\varepsilon^{3}\nonumber
\end{align}
so that the wall collisions would only be expected to be effective for a
vanishingly small fraction of the trajectories.

Turning now to the case $\omega_{r}\ll\omega_{o}$ (\ref{aaa}, Fig. 12 of
\cite{JMP}) we have as the condition that the suppression be effective:%

\begin{align*}
\beta &  =\frac{1}{\omega_{o}\tau_{c}}\geq1\\
\frac{1}{\omega_{o}}  &  \gtrsim\frac{2R}{v}\sin\alpha\\
\sin\alpha &  \leq\frac{v}{2R\omega_{o}}=\frac{200}{2\cdot25\cdot200}%
=\frac{1}{50}%
\end{align*}
for conditions typical of \cite{JMP} Fig. 12 ($v=200cm/\sec,B_{o}=1\mu T$).

Thus the wall collisions rate is never high enough to significantly effect the
magnitude of the frequency shift at the limits. The wall collisions do,
however, broaden and shift the resonances discussed in \cite{JMP} \bigskip

\section{Conclusion}

We have developed a general technique of analyzing the systematic effects due
to a combination of an electric field and magnetic gradients as encountered in
EDM experiments that employ gasses of stored particles. Use of the correlation
technique, either by numerical calculations for complicated geometries, or by
the velocity correlation function for simpler geometries, provides a
simplified approach to the problem compared to numerical integration of the
Bloch equations. Our analysis has added insight to this new systematic effect
and provides a means of rapidly assessing the effects of various geometries
and angular distributions for wall and gas collisions.

\section{Acknowledgements}

We are grateful to Werner Heil and Yuri Sobolev for calling our attention to
this problem and to George Jackeli and Boris Toperverg for an enlightening
conversation. We also thank J.M. Pendlebury et al. for providing a draft of
their manuscript before publication.

\section{Appendix 1: Matrix algebra of spherical Pauli matrices}

The following relationships among the Pauli matrices have been employed in the
calculation in section \ref{densmatr}.
\begin{equation}
2\sigma_{\pm}=\sigma_{x}\pm i\sigma_{y}%
\end{equation}%
\begin{equation}
\sigma_{\pm}\sigma_{z}=\mp\sigma_{\pm};\ \ \ \sigma_{z}\sigma_{\pm}%
=\sigma_{\pm}%
\end{equation}%
\begin{equation}
\sigma_{\pm}\sigma_{\mp}={\frac{1}{2}}\pm{\frac{1}{2}}\sigma_{z}%
\end{equation}%
\begin{equation}
\sigma_{z}\sigma_{z}=1;\ \ \ \sigma_{\pm}\sigma_{\pm}=0
\end{equation}

\section{Appendix 2: Egelstaff's velocity correlation function theorem; a new
look at an old theorem}

\subsection{Introduction}

The relation between the velocity autocorrelation function (vacf) and
$S_{s}\left(  q,\omega\right)  $ which we used in section [\ref{theorem}] was
first introduced by Egelstaff \cite{Egelstaff2} and has proven to be a useful
tool in the study of liquids. The vacf can be simulated for various models and
obtained from neutron scattering data using Egelstaff's theorem. The theorem
has been discussed by several authors \cite{Egelstaff}, \cite{Lovesey},
\cite{squires} and has been given in slightly different forms depending on the
normalization chosen for the functions involved.

Following Squires' \cite{squires} derivation would yield
\begin{equation}
\psi(\omega)=3\omega^{2}\lim_{q\rightarrow0}\frac{S_{inc}(q,\omega)}{q^{2}}
\label{33}%
\end{equation}
if we were to define
\[
\psi(\omega)=\frac{1}{2\pi}\int_{-\infty}^{\infty}\left\langle \overrightarrow
{v}(0)\cdot\overrightarrow{v}(\tau)\right\rangle e^{-i\omega\tau}d\tau
\]

which has a different normalization then used by Squires. Egelstaff gives the
theorem as
\[
\psi(\omega)=\omega^{2}\left[  \frac{S_{inc}(q,\omega)}{q^{2}}\right]
_{q\rightarrow0}%
\]
where he defines%
\[
\psi(\omega)=\frac{1}{2\pi}\int_{-\infty}^{\infty}\left\langle v_{x}(0)\cdot
v_{x}(\tau)\right\rangle e^{-i\omega\tau}d\tau
\]
which accounts for the factor of 3 difference. Both authors give the
derivation only for the Gaussian approximation where we take
\begin{align}
G(\overrightarrow{r},t)  &  =\frac{1}{\left(  2\pi w(\tau)\right)  ^{3/2}%
}e^{-\frac{^{r^{2}}}{2w(\tau)}}\nonumber\\
I(q,\tau)  &  =\int d^{3}rG(\overrightarrow{r},t)e^{i\overrightarrow{q}%
\cdot\overrightarrow{r}}=e^{-q^{2}\frac{w(\tau)}{2}} \label{3a}%
\end{align}

Boon and Yip \cite{boyip} derive the theorem for the general case, i.e.
without the Gaussian approximation. \qquad

The theorem has been used to extract vacf's from neutron scattering data by
many authors. An early example is given by \cite{Egelstaff2}. See also the
work of Carneiro \cite{carneiro}.

The fact that sections \ref{theorem} and \ref{notheorem} give the same result
and section \ref{theorem} uses Egelstaff's theorem while section
\ref{notheorem} does not, suggests that we have discovered a new way of
proving the theorem.

\subsection{A new derivation of Egelstaff's theorem}

In this section we will give a general derivation (not relying on the Gaussian
approximation) of Egelstaff's theorem.

We begin by following the formulation of Squires \cite{squires} and calculate
the velocity autocorrelation function%
\[
\psi(\tau)=\left\langle \overrightarrow{v}(0)\cdot\overrightarrow{v}%
(\tau)\right\rangle
\]
as follows:

Let $\overrightarrow{r}(t)$ be the position of a particle at time $t$, when
the particle was at the position $\overrightarrow{r}(0)$ at time $t=0$. Then
\begin{align*}
r(t)  &  \triangleq\overrightarrow{r}(t)-\overrightarrow{r}(0)=\int_{0}%
^{t}\overrightarrow{v}(t^{\prime})dt^{\prime}\\
r^{2}(t)  &  =\int_{0}^{t}\overrightarrow{v}(t^{\prime})dt^{\prime}\cdot
\int_{0}^{t}\overrightarrow{v}(t^{\prime\prime})dt^{\prime\prime}\\
&  =2\int_{0}^{t}dt^{\prime\prime}\int_{0}^{t^{\prime\prime}}\left\langle
\overrightarrow{v}(t^{\prime})\cdot\overrightarrow{v}(t^{\prime\prime
})\right\rangle dt^{\prime}%
\end{align*}
Now since $\left\langle \overrightarrow{v}(t^{\prime})\cdot\overrightarrow
{v}(t^{\prime\prime})\right\rangle =f(t^{\prime\prime}-t^{\prime})$ (for
stationary systems) we can write%
\[
r^{2}(t)=2\int_{0}^{t}\left\langle \overrightarrow{v}(0)\cdot\overrightarrow
{v}(t^{\prime})\right\rangle \left(  t-t^{\prime}\right)  dt^{\prime}%
\]
and%
\begin{align}
\frac{d}{dt}r^{2}(t)  &  =2\left\langle \overrightarrow{v}(0)\cdot
\overrightarrow{v}(t)\right\rangle t+2\int_{0}^{t}\left\langle \overrightarrow
{v}(0)\cdot\overrightarrow{v}(t^{\prime})\right\rangle dt^{\prime
}-2\left\langle \overrightarrow{v}(0)\cdot\overrightarrow{v}(t)\right\rangle
t\nonumber\\
&  =2\int_{0}^{t}\left\langle \overrightarrow{v}(0)\cdot\overrightarrow
{v}(t^{\prime})\right\rangle dt^{\prime}\nonumber\\
\frac{d^{2}}{dt^{2}}r^{2}(t)  &  =2\left\langle \overrightarrow{v}%
(0)\cdot\overrightarrow{v}(t)\right\rangle \label{a}%
\end{align}
Now, based on the usual definition of the pair distribution function,
$G(\overrightarrow{r},t)$, we have (following \cite{McGregor})
\begin{align*}
r^{2}(t)  &  =\int r^{2}G(\overrightarrow{r},t)d^{3}r\\
&  =\int r^{2}\frac{d^{3}r}{\left(  2\pi\right)  ^{3}}\int d^{3}%
qI(\overrightarrow{q},t)e^{i\overrightarrow{q}\cdot\overrightarrow{r}}\\
&  =\int d^{3}qI(\overrightarrow{q},t)\int\frac{d^{3}r}{\left(  2\pi\right)
^{3}}r^{2}e^{i\overrightarrow{q}\cdot\overrightarrow{r}}\\
&  =\int d^{3}qI(\overrightarrow{q},t)\int\frac{d^{3}r}{\left(  2\pi\right)
^{3}}\overrightarrow{r}\cdot\frac{1}{i}\overrightarrow{\triangledown}%
_{q}e^{i\overrightarrow{q}\cdot\overrightarrow{r}}\\
&  =\int d^{3}qI(\overrightarrow{q},t)\int\frac{d^{3}r}{\left(  2\pi\right)
^{3}}\frac{1}{i}\overrightarrow{\triangledown}_{q}\cdot\frac{1}{i}%
\overrightarrow{\triangledown}_{q}e^{i\overrightarrow{q}\cdot\overrightarrow
{r}}\\
&  =-\int d^{3}qI(\overrightarrow{q},t)\overrightarrow{\triangledown}_{q}%
\cdot\left(  \overrightarrow{\triangledown}_{q}\int\frac{d^{3}r}{\left(
2\pi\right)  ^{3}}e^{i\overrightarrow{q}\cdot\overrightarrow{r}}\right) \\
&  =-\int d^{3}qI(\overrightarrow{q},t)\overrightarrow{\triangledown}_{q}%
\cdot\left(  \overrightarrow{\triangledown}_{q}\delta^{3}\left(
\overrightarrow{q}\right)  \right)
\end{align*}
where we introduced the spatial Fourier transform of the pair distribution
function, $I(\overrightarrow{q},t).$ (See the first part of equation \ref{3a})

Integrating by parts twice we obtain%
\begin{align*}
r^{2}(t)  &  =\int d^{3}q\overrightarrow{\triangledown}_{q}I(q,t)\cdot\left(
\overrightarrow{\triangledown}_{q}\delta^{3}\left(  \overrightarrow{q}\right)
\right) \\
&  =-\int d^{3}q\delta^{3}\left(  \overrightarrow{q}\right)  \overrightarrow
{\triangledown}_{q}\cdot\left(  \overrightarrow{\triangledown}_{q}%
I(q,t)\right) \\
&  =-\lim_{q\rightarrow0}\overrightarrow{\triangledown}_{q}\cdot\left(
\overrightarrow{\triangledown}_{q}I(q,t)\right)  =-\lim_{q\rightarrow0}\left(
\triangledown_{q}^{2}I(q,t)\right)
\end{align*}
Then, using (\ref{a}) we have%
\begin{equation}
\psi\left(  \tau\right)  =\left\langle \overrightarrow{v}(0)\cdot
\overrightarrow{v}(\tau)\right\rangle =-\frac{1}{2}\lim_{q\rightarrow0}\left(
\triangledown_{q}^{2}\frac{d^{2}}{d\tau^{2}}I(q,\tau)\right)  \label{b}%
\end{equation}

Writing%
\begin{align*}
\psi\left(  \tau\right)   &  =\int_{-\infty}^{\infty}e^{i\omega\tau}%
\psi\left(  \omega\right)  d\omega\\
I(q,\tau)  &  =\int_{-\infty}^{\infty}e^{i\omega\tau}S\left(  q,\omega\right)
d\omega
\end{align*}
we find (in general)
\begin{equation}
\psi\left(  \omega\right)  =\frac{\omega^{2}}{2}\lim_{q\rightarrow0}\left(
\triangledown_{q}^{2}S(q,\omega)\right)  \label{d}%
\end{equation}

This appears to be different than the usual form of the theorem (\ref{33}) but
is completely equivalent as can be seen by using%

\[
S_{s}\left(  q,\omega\right)  =\int dte^{-i\omega t}\left\langle
e^{i\overrightarrow{q}\cdot\left(  \overrightarrow{R}(t)-\overrightarrow
{R}(0)\right)  }\right\rangle
\]
and expanding for small $q$.%
\[
S_{s}\left(  q,\omega\right)  =\int dte^{-i\omega t}\left\langle
1+i\overrightarrow{q}\cdot\left(  \overrightarrow{R}(t)-\overrightarrow
{R}(0)\right)  -\frac{\left[  \overrightarrow{q}\cdot\left(  \overrightarrow
{R}(t)-\overrightarrow{R}(0)\right)  \right]  }{2}^{2}\right\rangle
\]
The first term gives a $\delta\left(  \omega\right)  $ which does not
contribute to the result, the second term averages to zero and the third term
gives
\[
S_{s}\left(  q,\omega\right)  =\alpha q^{2}%
\]
(for isotropic media). Then%
\[
\lim_{q\rightarrow0}\left(  \triangledown_{q}^{2}S(q,\omega)\right)  =6\alpha
\]
and equation (\ref{d}) is equivalent to (\ref{33}) in general.

For the Gaussian approximation%
\[
I(q,\tau)=e^{-q^{2}\frac{w(\tau)}{2}}%
\]
we have%
\[
\lim_{q\rightarrow0}\frac{d^{2}}{d\tau^{2}}I(q,\tau)=-\frac{q^{2}}%
{2}\frac{d^{2}w}{d\tau^{2}}%
\]
and using (\ref{b})%
\begin{equation}
\psi\left(  \tau\right)  =-\frac{1}{2}\lim_{q\rightarrow0}\left(
\triangledown_{q}^{2}\frac{d^{2}}{d\tau^{2}}I(q,\tau)\right)  =\frac{3}%
{2}\frac{d^{2}w}{d\tau^{2}}=-\lim_{q\rightarrow0}\frac{3}{q^{2}}\frac{d^{2}%
}{d\tau^{2}}I(q,\tau) \label{c}%
\end{equation}

Finally
\begin{equation}
\psi\left(  \omega\right)  =\lim_{q\rightarrow0}\frac{3}{q^{2}}\omega
^{2}S(q,\omega) \label{e}%
\end{equation}
for the Gaussian approximation.(note that $\nabla_{q}^{2}q^{2}=6$), confirming
that our formulation again gives the correct result.

\subsection{Discussion}

For the case of diffusion, at very long times the Gaussian approximation holds
$\left(  w(\tau\right)  =2D\tau)$, but $d^{2}w/d\tau^{2}=0$, so that the
calculation in (\ref{c}) appears to break down. Nonetheless the more general
derivations show that one can apply Egelstaff's theorem to this case where
\begin{equation}
S(q,\omega)=\frac{1}{\pi}\frac{Dq^{2}}{\omega^{2}+\left(  Dq^{2}\right)  ^{2}%
}.
\end{equation}
as was done in section \ref{theorem}.

\bigskip

\vfill  \eject

\section{\bigskip Figures}%

\begin{figure}
[h]
\begin{center}
\includegraphics[
width=3.8657in
]%
{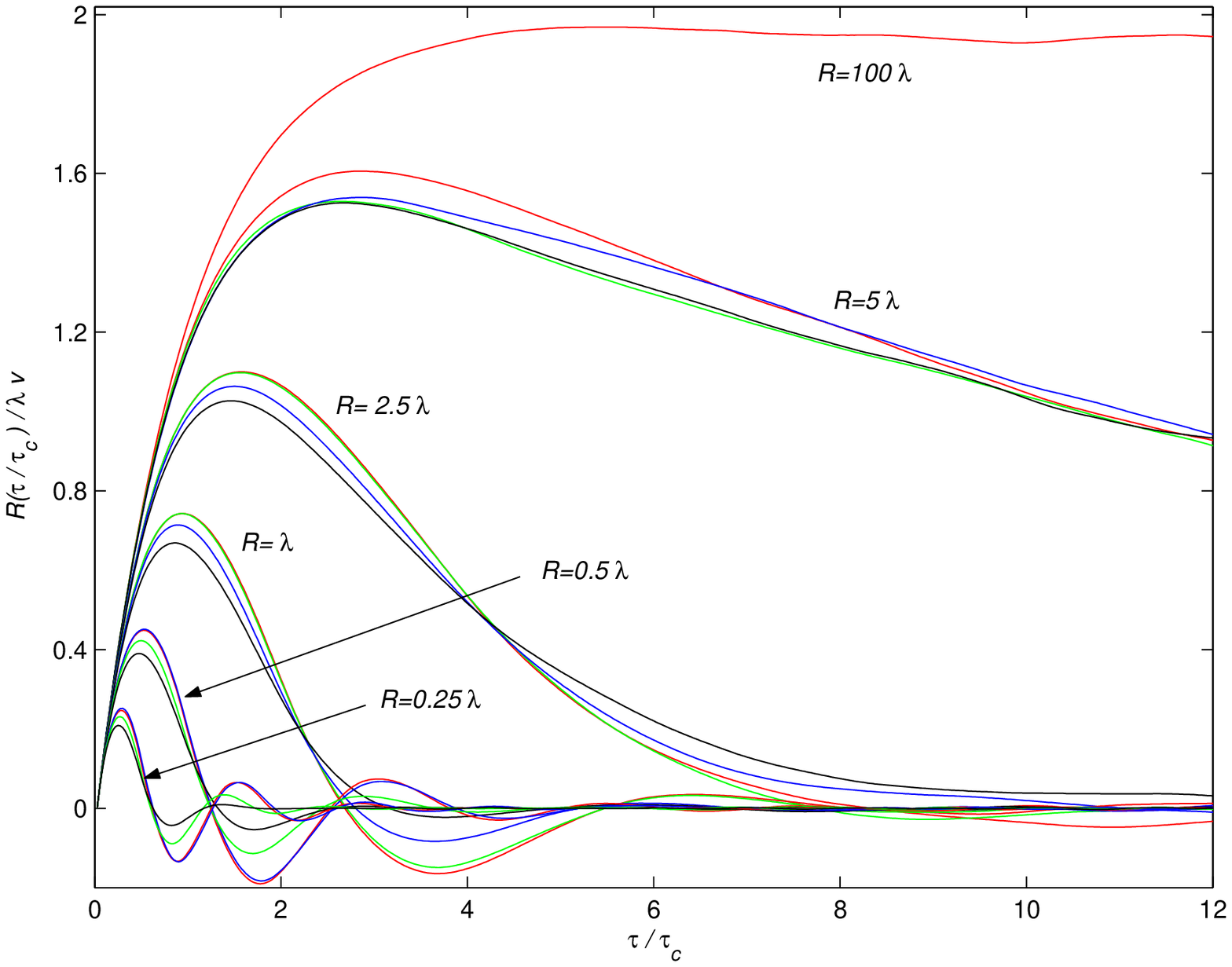}%
\caption{The position-velocity correlation function, $R\left(  \tau\right)
=2h\left(  \tau\right)  $ as a function of cell radius $R$ parameterized in
terms of the mean free path $\lambda$ for different degrees of specularity as
parameterized by the angular spread of the final angle compared to the
incident angle. Red: specular; Green: $45^{\circ}$; Blue: $90^{\circ}$; Black:
diffuse. For $R\gtrsim2.5\lambda$, there was practically no effect due to the
degree of specularity, as expected.}%
\label{11}%
\end{center}
\end{figure}

\begin{figure}
[h]
\begin{center}
\includegraphics[
width=3.8917in
]%
{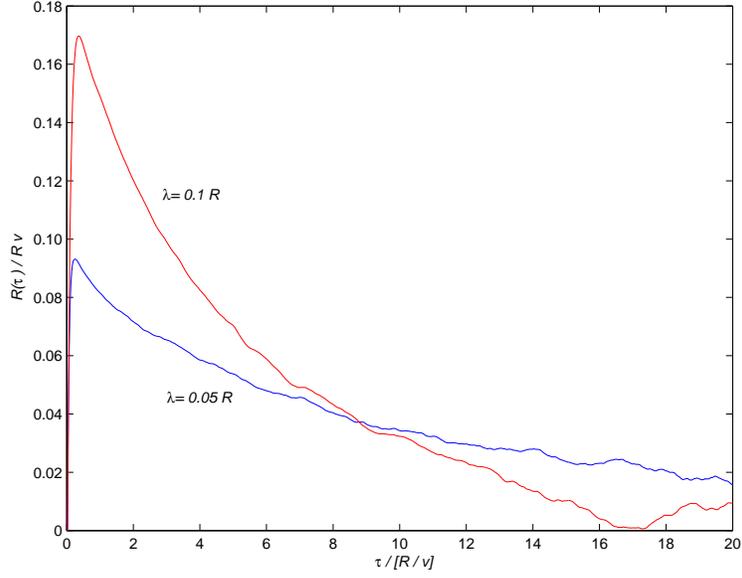}%
\caption{The position-velocity correlation function when $\lambda$ is very
small. This represents the limit for slow diffusion.}%
\end{center}
\end{figure}

\begin{figure}
[h]
\begin{center}
\includegraphics[
width=3.9237in
]%
{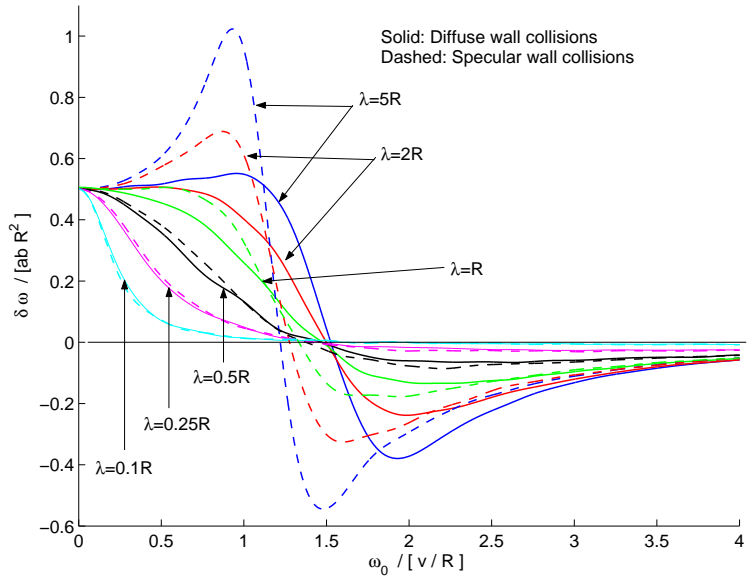}%
\caption{Results of numerically applying Eq. (\ref{ab1}) to numerical
calculations of the correlation function, for varying $\lambda$ with $R$
fixed.}%
\end{center}
\end{figure}

\bigskip%
\begin{figure}
[h]
\begin{center}
\includegraphics[
width=4.868in
]%
{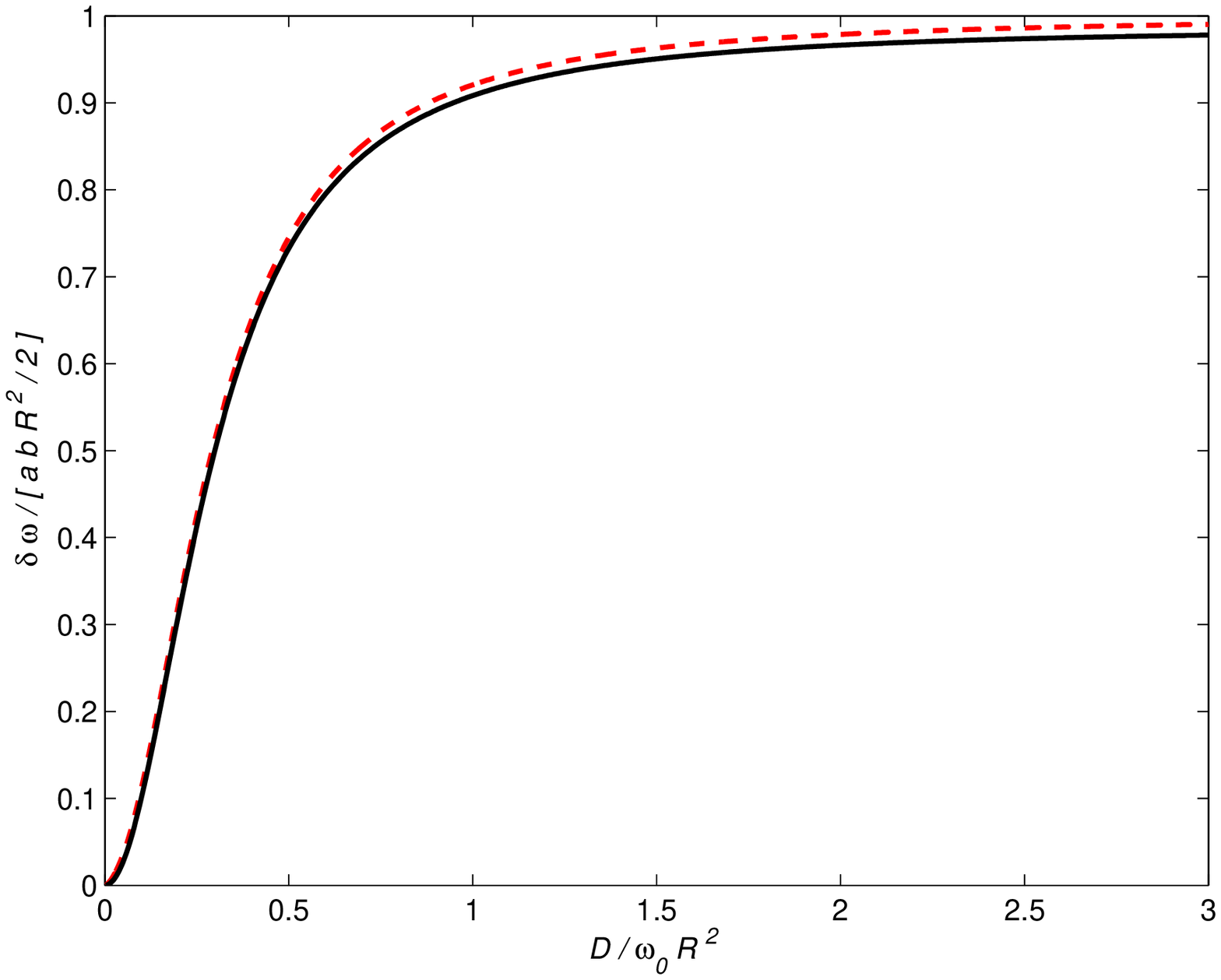}%
\caption{$2\delta\omega/abR^{2}$ versus $y={D}/{R^2\omega_{o}}$,
equation (\ref{big}). Black first term, red sum of first 4 terms}%
\end{center}
\end{figure}

\begin{thebibliography}{9}
\bibitem {degenn}This theorem, apparently based on work of DeGennes [12], was
first stated by Egelstaff [13], and has been cited by many authors, (e.g.
[13,14,15]), with slightly different numerical constants. The most complete
proof is given in [15]. The form we have used is appropriate for our \ two
dimensional problem. In appendix 2 we give a detailed discussion and a new
proof of the theorem.\
\end{thebibliography}
\end{document}